\documentclass{iopart_mod}

\usepackage{amsmath}
\usepackage{amssymb}
\usepackage{epsfig}

\DeclareMathOperator{\extdm}{d}
\newcommand{\extd}{\extdm \!}

\newcommand{\beqas}{\begin{eqnarray*}}
\newcommand{\beqa}{\begin{eqnarray}}

\newcommand{\eeqas}{\end{eqnarray*}}
\newcommand{\eeqa}{\end{eqnarray}}

\newcommand{\eps}{\varepsilon}
\newcommand{\al}{\alpha}

\newcommand{\ga}{\gamma}

\newcommand{\la}{\lambda}

\newcommand{\Ga}{\Gamma}

\newcommand{\Om}{\Omega}

\newcommand{\blist}{\begin{itemize}}

\newcommand{\elist}{\end{itemize}}

\providecommand{\href}[2]{#2}

\DeclareFontFamily{OT1}{rsfs}{}
\DeclareFontShape{OT1}{rsfs}{m}{n}{ <-7> rsfs5 <7-10> rsfs7 <10->rsfs10}{} 
\DeclareMathAlphabet{\mycal}{OT1}{rsfs}{m}{n}

\newcommand{\coup}{N}

\begin{document}

%\hfill LU-ITP--05--xxx

\title{The Volume of 2D Black Holes}

\author{Daniel Grumiller}
 
\address{Institut f\"ur
    Theoretische Physik, Universit\"at Leipzig, \\ Augustusplatz 10-11,
    D-04109 Leipzig, Germany}

\ead{grumil@hep.itp.tuwien.ac.at}

\pacs{04.70.-s, 04.60.Kz, 11.25.Pm, 97.60.Lf}

%\submitto{Journal of Physics: Conference Series}

\begin{abstract}
  
It is shown that the definition for the volume of stationary black holes advocated in \cite{Parikh:2005} readily generalizes to the case of dilaton gravity in $D=2$. The dilaton field is included as part of the measure. A feature observed in $D=3$ and $4$ has been the impossibility to obtain infinite volume while retaining finite area without encountering some kind of pathology. It is demonstrated that this also holds in $D=2$. Consistency with spherically reduced gravity is shown. For the Witten black hole it is found that the area is proportional to the volume. %Volume and area of the exact string black hole are discussed in some detail.
 
\end{abstract}

\setcounter{footnote}{0}
\section{Introduction}

In General Relativity the volume of some region in space depends on how spacetime is sliced into a spatial part and time. This is unsatisfactory insofar as it appears to prohibit a meaningful definition of the volume of black holes (BHs) and thus to quantify the ratio between the entropy as expected naively from quantum field theory -- which grows with the volume because entropy is an extensive quantity -- and the Bekenstein-Hawking entropy as derived from the laws of BH mechanics -- which grows with the area of the BH.  

In Ref.~\cite{Parikh:2005} M.~Parikh has provided a slicing-invariant definition for the volume of stationary black holes. Moreover, he could show that for dimension $D=3$ and $4$ the volume can never be infinite if the area is finite, which led to interesting possible implications for the Bekenstein-Hawking entropy. It has been pointed out as well that in $D>4$ there may be loopholes which, in principle, could allow to circumvent these nogo results. The case $D=2$ has not been addressed.

It is the purpose of this work to extend such a discussion to $D=2$. Section \ref{se:2} provides the definition and treats several examples. In section \ref{se:3} the question is addressed whether a BH with finite area might have infinite volume. Section \ref{se:4} concludes with a brief discussion. The \ref{app:A} contains an application to the exact string BH.

%In the last section the definition is applied to spherically reduced gravity, where consistency with the fourdimensional case is obtained, and to the Witten BH, where the volume is found to be proportional to the area of the BH. Implications for scalar-tensor theories in higher dimensions are addressed.

\section{The volume of a 2D black hole}\label{se:2}
 
Stationary BHs in $D=2$ emerge as classical solutions of 2D dilaton gravity \cite{Grumiller:2002nm}, the action of which is given by\footnote{As the Einstein-Hilbert action in 2D yields no equations of motion it is not a sensible starting point for model building. The next natural thing to consider are scalar-tensor theories. The units $16\pi G_N=\hbar=c=1=k_B$ are used.}
\begin{equation}
  \label{eq:action}
  L=\coup\int_{\mathcal{M}_2}\extd^2x\sqrt{-g}\left[XR-U(X)(\nabla X)^2+\la^2 \tilde{V}(X)\right]\,.
\end{equation}
The curvature scalar $R$ and covariant derivative $\nabla$ are associated with the Levi-Civit\'a connection related to the metric $g_{\mu\nu}$, the determinant of which is denoted by $g$. The dimensionless scalar field $X$ often is called ``dilaton field''. In the present work semi-positivity will be assumed, $X\geq 0$. To avoid confusion it should be noted that in string literature almost exclusively the dilaton field $\phi$ defined by $X=e^{-2\phi}$ is used instead of $X$.
The dimensionless functions $U,\tilde{V}$ define the model, which may or may not exhibit BH solutions. The scaling constant $\la$ with length dimension minus one will play no role in the present work.
The numerical overall constant $\coup$ will be kept for the time being. 
It is useful to define 
\begin{equation}
  \label{eq:if}
  Q(X):=\int^X\extd y\, U(y)\,.
\end{equation}
The additive ambiguity inherent to this definition may be fixed conveniently. Physically it corresponds to the freedom to choose the unit of mass.

All classical solutions of \eqref{eq:action} yield a line element that may be brought into the following form:
\begin{equation}
  \label{eq:le}
  \extd s^2=\al(r)\extd t^2+2\al(r)f'\extd t \extd r-\left(\frac{1}{\al(r)}-\al(r) {f'}^2\right)\extd r^2
\end{equation}
Prime denotes differentiation with respect to $r$.
By inspection, there is always a Killing vector $k^\mu\partial_\mu=\partial_t$.
Up to different notations for signature this coincides with the $t-r$ part of %Eq. (6) 
the line element considered in Ref.~\cite{Parikh:2005}. Different choices of the function $f$ correspond to different time-slicings.
The transformation $\extd u=\extd t+(f'-1/\al)\extd r$, $\extd r = \exp{Q(X)}\extd X$ puts \eqref{eq:le} into Eddington-Finkelstein gauge,
\begin{equation}
  \label{eq:EF}
  \extd s^2=2e^{Q(X)}\extd u\left(\extd X - (M+w(X))\extd u\right)\,.
\end{equation}
This coincides with the general solution of \eqref{eq:action}, Eq.~(3.24) of \cite{Grumiller:2002nm} (up to minor notational differences; cf.~also \cite{Klosch:1995fi}). The constant of motion appearing in \eqref{eq:EF}, denoted by $M$, is related to the ADM mass (whenever this notion makes sense). The function 
\[
w(X):=\frac{\la^2}{2}\int^X\extd y \,e^{Q(y)}\tilde{V}(y)
\]
contains an additive ambiguity which may be fixed, for instance, by defining the ground state solution $M=0$ to be Minkowski space or a BPS solution (whenever such solutions exist). The function $\al(r)$ in \eqref{eq:le} is proportional to $(M+w(X))\exp{Q(X)}$. Thus, Killing horizons are determined either by zeros of $\al(r)$ or by the equation $M+w(X)=0$. Henceforth exclusively the line element \eqref{eq:EF} will be enga(u)ged.

The determinant of the spacetime metric has no dependence on the time slicing,
\begin{equation}
  \label{eq:detg}
  \sqrt{-g(X)}=e^{Q(X)}\,,
\end{equation}
and thus analogy to \cite{Parikh:2005} suggests a volume definition of the form $V:=\int\extd r\sqrt{-g(r)}=\int\extd X\exp{Q(X)}$. However, one can easily see that such a volume definition would be inappropriate for thermodynamical considerations, which have been a key motivation in \cite{Parikh:2005}: the Euclidean action -- and thus also free energy, the partition function and entropy -- scale with the dilaton field because the curvature scalar in the action \eqref{eq:action} is multiplied by $X$. Also the Gibbons-Hawking boundary term is multiplied under the integrand by this factor. Therefore, one ought to consider the dilaton field as being part of the measure. Before providing our proposal for the definition of the volume the one for the area is recalled:
\begin{equation}
  \label{eq:vol4}
  A:= N X\,.
\end{equation}
Again, the reasoning is the same as above: area, which in 2D gravity actually is zero-dimensional, should scale with the dilaton field in order to be consistent with the Bekenstein-Hawking relation $S=A_h/(4G_N)$, where $S$ is the thermodynamical entropy of a 2D BH and $A_h$ the area evaluated at the BH horizon. Entropy has been calculated in \cite{Gegenberg:1995pv} by simple thermodynamical considerations invoking the first law and also by Wald's Noether charge technique. Translated to our notation the result is\footnote{The normalization $N$ in \eqref{eq:action} is divided by $16\pi G_N$, which in our convention equals unity. As a consequence $S=4\pi A$.}
\begin{equation}
  \label{eq:entropy}
  S=4\pi N X_h\,.
\end{equation}
As a consistency check one may consider the Schwarzschild BH (SBH). Because $N=4\pi$ is just the volume of the angular part that has been integrated out to obtain \eqref{eq:action} and $X=r^2$, the normalization in \eqref{eq:vol4} is seen to be correct. 
For dimensionally reduced models the emergence of the dilaton field in the measure is not surprising.\footnote{For instance, spherically symmetric Einstein gravity in $D=4$ leads to a line element $\extd s^2=g_{\mu\nu}^{(2)}(x^\ga)\extd x^\mu\extd x^\nu - X(x^\ga)\extd\Om^2_{S^2}$, the determinant of which is determined by $\extd^4x\sqrt{-g^{(4)}}=\extd^2\Om_{S^2}\extd^2x\sqrt{-g^{(2)}}|X|$.}
But also from an intrinsically 2D point of view the appearance of the dilaton field in \eqref{eq:vol4} is not unexpected if one keeps in mind that $X$ is essentially the (space-time dependent) inverse Newton coupling.

Thus, the volume in 2D dilaton gravity will be defined by
\begin{equation}
  \label{eq:vol1}
  V:=c\int\extd r \sqrt{-g(r)} \,X(r)  \,.
\end{equation}
The overall normalization $c$ will be chosen later conveniently. One may absorb $c$ in the integration constant of \eqref{eq:if}, but we will refrain from doing so and assume the latter already being fixed by selecting a mass scale. For BHs we have to integrate from some lower value to the horizon. Inserting the adapted coordinates \eqref{eq:EF} establishes
\begin{equation}
  \label{eq:vol1.5}
 V=c\int\limits^{X_h}_{X_0} \extd X e^{Q(X)}X\,. 
\end{equation}
The quantity $X_h$ is the dilaton field evaluated at the Killing horizon 
(defined by $M+w(X)=0$), while $X_0$ is the lowest possible value the dilaton field may take inside the horizon. Typically, $X=X_0$ will coincide with the locus of a curvature singularity. In many cases $X_0=0$ is valid. 

For a large class of models $\exp{Q}=X^{-a}$ holds, in particular for the so-called $ab$-family of models (cf.~Ref.~\cite{Katanaev:1997ni} and Section 3.3 in \cite{Grumiller:2002nm}),
\begin{equation}
  \label{eq:vol5}
  U(X)=-\frac{a}{X}\,,\quad \tilde{V}\propto X^{a+b}\,.
\end{equation}
Inserting this into the definition yields
\begin{equation}
  \label{eq:vol2}
  V = \frac{c}{2-a} \left(X^{2-a}-X^{2-a}_0\right)\,,\quad\quad {\rm if\,\,} a\neq 2\,,
\end{equation}
and
\begin{equation}
  \label{eq:vol3}
  V = c\ln{\frac{X}{X_0}}\,,\quad\quad {\rm if\,\,} a=2\,.
\end{equation}
Note that the last case is very special insofar as neither $X_0=0$ nor $X_0=\infty$ yield a finite volume, irrespective of the area. This issue will be addressed in section \ref{se:3}. 
Let us now study some examples explicitly (always setting $X_0=0$).

\paragraph{Schwarzschild BH}
First of all the spherically reduced SBH, $a=1/2$, will be treated. In this case the dilaton field is essentially the surface area. By adjusting $\la$ in \eqref{eq:action} appropriately one may achieve $X=r^2$, where $r$ is the surface radius. If $c:=N/2=2\pi$ plugging these values into \eqref{eq:vol2} gives
\begin{equation}
  \label{eq:SSBH}
  V(\rm SBH) = \frac{4\pi}{3} r^3_h\,.
\end{equation}
This coincides with %Eq.~(13) 
the corresponding equation in \cite{Parikh:2005}. It is to be noted that without the dilaton field in the definition \eqref{eq:vol1} it would have been impossible to obtain \eqref{eq:SSBH}. 

%\paragraph{Reissner-Nordstr\"om BH} For BHs with two horizons it seems adequate to calculate the volume between them, e.g.~$r_{h}^{(1,2)}=M\pm\sqrt{M^2-Q^2}$ for the Reissner-Nordstr\"om BH. This is straightforward by virtue of \eqref{eq:SSBH}. For a near extremal BH, $M=|Q|+\eps$, one obtains $V\approx 2\sqrt{2Q\eps}A$, with $A=4\pi Q^2$.

\paragraph{Spherically reduced gravity for generic $D$}
Integrating out the angular part in the $D$-dimensional Einstein-Hilbert action yields a 2D dilaton gravity model which belongs to the $ab$-family, with $a=(D-3)/(D-2)$. Thus, the volume reads
\begin{equation}
  \label{eq:vol6}
  V(D) = c\,\frac{D-2}{D-1} X^{(D-1)/(D-2)}_h\,.
\end{equation}
Recalling that the surface radius is related to the dilaton via $X=r^{D-2}$ and that the properly normalized action implies 
\begin{equation}
  \label{eq:vol7}
  N=\frac{2\pi^{(D-1)/2}}{\Ga\left(\frac{D-1}{2}\right)}\,,
\end{equation}
the BH area, Eq.~\eqref{eq:vol4}, is seen to be correct. If one fixes
\begin{equation}
  \label{eq:vol8}
  c=N(1-a)=\frac{N}{D-2}\,,
\end{equation}
the ratio between volume and area yields
\begin{equation}
  \label{eq:vol9}
  \frac{V}{A}=\frac{1}{D-1}X_h^{-1+(D-1)/(D-2)} = \frac{r_h}{D-1}\,. 
\end{equation}
This demonstrates consistency with the result in the original dimension. 
%The reason for this consistency is that the dilaton field encodes information of the higherdimensional theory. Thus, despite of the intrinsically $D=2$ treatment the correct higherdimensional formulas are reproduced, provided the dilaton field is included in the measure according to \eqref{eq:vol1} and $c$ is chosen appropriately. 

\paragraph{Models with $U=0$}
If $a=0$ the volume is $V=cX^2_h/2$. Thus, the volume scales quadratically with the area and hence with entropy. The most prominent member belonging to this class is the Jackiw-Teitelboim model \cite{JT}, the solutions of which are %either Minkowski space or 
$(A)dS_2$.

\paragraph{Witten BH}
Finally, we address the Witten BH \cite{Witten:1991yr}, 
which has $a=1$. From \eqref{eq:vol2} and \eqref{eq:vol4} the ratio
\begin{equation}
  \label{eq:vol11}
  \left.\frac{V}{A}\right|_{\rm Witten\,\,BH} = \frac{c}{N}
\end{equation}
is seen to be independent from the dilaton field. Thus, for the Witten BH the area is proportional to the volume of the BH, with some fixed proportionality constant. It may be set to unity by choosing $c=N$.

%It is emphasized that the choice \eqref{eq:vol8} appears to be suitable only for spherically reduced models. Generically there does not seem to be a way to fix $c$ canonically; of course, for simplicity one may set it to unity for models which do not follow from dimensional reduction. In fact, for the $ab$-family \eqref{eq:vol5} by a redefinition of the dilaton field one may set $c=1$ if simultaneously $\la$ is rescaled in the action. This is possible because the first two terms in \eqref{eq:action} are homogeneous of degree one in the dilaton field and because $\tilde{V}$ is monomial.

\paragraph{Concluding remarks} Conformal transformations are a useful tool, especially in 2D. They do not leave invariant the functions $Q$ and $U$. Indeed, it is always possible to eliminate the kinetic term in \eqref{eq:action} by transforming to a different, conformally related, model. Thus, also the definition of the volume \eqref{eq:vol1} obviously depends on the choice of the conformal frame. 

\section{Finite area but infinite volume?}\label{se:3}

We ask now the question: is it possible to obtain an infinite volume while keeping the area bounded from above? Obviously, for finite values of $X_h$ this is impossible. On the other hand, $X_h\to\infty$ would blow up the area. Thus, the only candidate is the limit $X_h\to 0$. As can be seen from \eqref{eq:vol2} for $a<2$ also the volume shrinks to zero. However, for $a\geq 2$ the volume diverges. Is it therefore possible in $D=2$ to obtain infinite volume with zero area without encountering any pathologies?

The answer is no. To understand this one has to consider the global structure carefully.\footnote{For $a>2$ there is a non-rigorous shortcut: the definition \eqref{eq:vol2} suggests to take $X_0=\infty$ rather than $X_0=0$. So the ``outside observer'' would be located near $X=0$ rather than near $X=\infty$. But then it is not surprising that the limit $X\to 0$ yields an infinite volume -- after all, the volume calculated in this manner is the {\em total} volume of spacetime. However, this scenario would imply that the area of the ``asymptotic region'' vanishes rather than diverging. From a physical point of view this is pathological, as the area surrounding a BH should not be smaller than the one of the BH. The method in the text has the advantage of being rigorous and being applicable also to the limiting case $a=2$. We will assume henceforth that the asymptotic region is at $X=\infty$. The lower boundary $X_0$ will be left unspecified.} Again, we restrict ourselves to the $ab$-family \eqref{eq:vol5} for sake of conciseness and comment on the general case afterwards. The curvature scalar reads \cite{Grumiller:2002nm,Katanaev:1997ni}
\begin{equation}
  \label{eq:curvab}
  R=c_1a M X^{a-2} + c_2\frac{b(b+1-a)}{b+1}X^{a+b-1}\,,
\end{equation}
where $c_1,c_2$ are some non-vanishing normalization constants and $b\neq -1$.
\begin{figure}
\epsfig{file=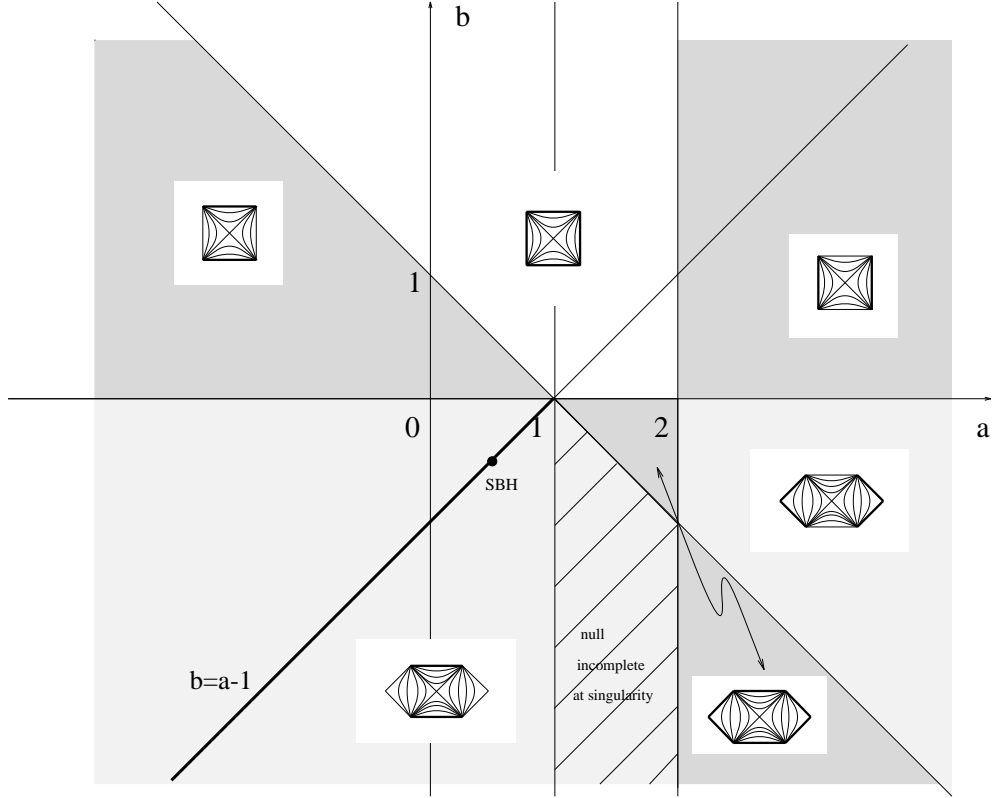,width=\linewidth}
\caption{Global structure of $ab$-family. Bold lines in the Carter-Penrose diagrams denote geodesically incomplete boundaries, curved lines are $X=\rm const.$ slices and the two diagonals denote the Killing horizon, intersecting at the bifurcation point. This figure is taken from Refs.~\cite{Grumiller:2002nm,Katanaev:1997ni}.}
\label{fig:1}
\end{figure}
As can be seen from Fig.~\ref{fig:1} for $a>2$ what we have considered so far as ``asymptotic region'', namely $X\to\infty$, is geodesically incomplete, thus becoming a singularity rather than an asymptotic region (at the same time, for $b>1-a$ the ``inner region'' in the corresponding Carter-Penrose diagram has a geodesically complete boundary). Thus, as might have been anticipated, these solutions are pathological, the pathology being the geodesically incomplete ``asymptotic region''. 
By the same token also the half line $b=-1, a>2$ may be excluded.

What remains to be discussed is the limiting case $a=2$ (we recall the logarithmic scaling according to \eqref{eq:vol3}). If $b\neq -1,0,1$ then the ground state solution ($M=0$ in \eqref{eq:EF}) according to \eqref{eq:curvab} has a naked singularity either at $X=0$ (if $b<-1$) or at $X=\infty$ (if $b>-1$); thus, models with $a=2$ and $b\neq -1,0,1$ have no regular ground state solution to which the BH may evaporate, a property which should be considered as pathological. If $b=-1$ the curvature scalar turns out to contain a constant term and a term scaling logarithmically with the dilaton field. This implies two curvature singularities, one at $X=0$ and another one at $X=\infty$. Thus, the asymptotic region is singular.
The isolated models $a=2$ and $b=0$ or $b=1$ are somewhat recalcitrant. Apparently, there is no singularity because the curvature scalar \eqref{eq:curvab} is just a constant proportional to $M$. So we seem to have two counter examples to the claim above.  However, the standard definition of singularities in General Relativity invokes not some invariants of the Riemann tensor but rather (in)completeness properties of geodesics. They have been studied exhaustively for the $ab$-family in Ref.~\cite{Katanaev:1997ni}. For both cases the boundary $X=\infty$ turns out to be incomplete with respect to null and non-null geodesics. Thus, the ``asymptotic region'' is singular, despite of curvature remaining bounded (in fact, constant). Again a pathology arises.\footnote{For $b=1$ one may nevertheless provide an appropriate volume definition in a different conformal frame, namely $a=0$. This is nothing but the Jackiw-Teitelboim model \cite{JT} which has also constant curvature solutions. Thus, the conformal transformation between these two models leaves curvature intact, which is quite remarkable (although there is actually a difference: for the Jackiw-Teitelboim model the constant curvature is determined by $c_2$ in \eqref{eq:curvab} and thus the same for all solutions, while for $a=2$, $b=1$ curvature is given by $2c_1M$ and thus differs from solution to solution; so although individual solutions may be mapped onto each other, there is no mapping between the families of solutions). However, because of the singularity of the conformal factor (essentially $X^2$) completeness properties and the causal structure may change, which is indeed what happens: the solutions of the Jackiw-Teitelboim model are geodesically complete at $X=\infty$. On a related note, the model $a=2$, $b=0$ is conformally related to the Witten BH \cite{Witten:1991yr}; again the corresponding conformal factor exhibits two singularities. These features indicate that for each choice of $b$ there should be some ``preferred'' conformal frame in which the volume ought to be calculated; it may depend on the particular application which frame this is, but quite often $a=1+b$ is the best choice, because only in this frame the ground state solution ($M=0$) is Minkowski space. Fig.~\ref{fig:1} suggests to take canonically $a=1-|b|$ (with $b$ given) for the ``best behaved'' Carter-Penrose diagram.}

Actually, similar arguments and inspection of the curvature scalar \eqref{eq:curvab} establish that only the following class of $ab$-models reveal none of these pathological features: %\footnote{When these inequalities are saturated, i.e., $a=1$ and $b=0$, then the Witten BH \cite{Witten:1991yr} is encountered which exhibits also a feature that might be considered as pathological: the singularity is null complete \cite{Katanaev:1997ni}. However, one should not take this geometry at face value close to the singularity and rather work with the exact string BH \cite{Dijkgraaf:1992ba} which has no singularity and asymptotes to the Witten BH for large values of $X$.} 
$a=1+b\leq 1$ (Minkowskian ground state models, to which spherically reduced gravity theories belong), $a=1-b\leq 1$ ($(A)dS$ ground state models, to which the Jackiw-Teitelboim model \cite{JT} belongs) and $b=0,a\leq 1$ (Rindler ground state models, to which the Witten BH \cite{Witten:1991yr} belongs). These special lines in the $ab$-plane have been depicted in Fig.~\ref{fig:1}. 

We conclude this section by remarking that the same arguments apply even if $\tilde{V}$ is an arbitrary polynomial as long as $\exp{Q}=X^{-a}$, because 1.~the volume definition \eqref{eq:vol2} is independent from $\tilde{V}$ and 2.~for large/small values of $X$ such a model will essentially belong to the $ab$-family, so the discussion above may be employed with appropriate refinements. With a few exceptions these considerations cover practically all 2D dilaton gravity models of interest, and it may be expected that even for those exceptions (cf.~\ref{app:A} for a particular example of relevance) a similar mechanism exists which renders the volume either finite or the global structure pathological.

\section{Discussion}\label{se:4}

We have seen that both of the key results of \cite{Parikh:2005} readily generalize to BHs in $D=2$: there is an analogous definition of the volume of BHs, Eq.~\eqref{eq:vol1}, and it is impossible to encounter an infinite volume in a non-singular manner while keeping the area finite. Just as in $D=3,4$ there is a mechanism that prevents the volume from diverging for bounded values of the area, which may be taken as a further indication that this is a generic property. 

Consistency with spherically reduced gravity from any dimension has been shown and a peculiar feature of the Witten BH has been pointed out (which also holds for the exact string BH for large masses): its area is proportional to the volume, with some universal proportionality constant.  This may have interesting implications for the interpretation of the Bekenstein-Hawking entropy of this string-inspired BH solution.

Finally, it is suggestive to propose a definition analogous to \eqref{eq:vol1} for higherdimensional scalar-tensor theories in the Jordan frame/string frame. This could help to shed additional light on the issue of whether a stationary BH with finite area might have infinite volume.

\ack

I am grateful to M.~Parikh for helpful correspondence. I thank the organizers, M.~Cadoni, M.~Cavagli\`a and J.E.~Nelson, and the participants of the conference ``Fourth Meeting on Constrained Dynamics and Quantum Gravity'' for the enjoyable atmosphere. This work has been supported by project J2330-N08 of the Austrian Science Foundation (FWF).

\appendix

\newcommand{\arctanh}{{\rm arctanh}\,}
\newcommand{\arcsinh}{{\rm arcsinh}\,}

\section{Application to the exact string black hole}\label{app:A}

As a particular counter example to models with $\exp{Q}=X^{-a}$ the exact string BH (ESBH) \cite{Dijkgraaf:1992ba} will be discussed in this appendix. In the whole range of definition its volume will turn out to be finite.

For the ESBH recently a unique target space action could be constructed, which allowed for the first time a calculation of its mass and entropy \cite{Grumiller:2005sq}. It turned out to be an action of type \eqref{eq:action} with an additional abelian $BF$-term, which is irrelevant for the present context. The potentials 
\begin{equation}
  \label{eq:solutionofESBH3}
  \tilde{V}=-2\gamma\,,\qquad U= -\frac{1}{\gamma + 2\left(\frac{1}{\gamma}+\sqrt{1+\frac{1}{\gamma^2}}\right)}\,.
\end{equation}
turned out to be rather simple functions of an auxiliary field $\ga$, related to the relevant dilaton field $X$ via
\begin{equation}
  \label{eq:solutionofESBH2.5}
  X = \gamma + \arcsinh{\gamma}\,.
\end{equation}
Only the first term in \eqref{eq:solutionofESBH2.5} is anticipated from a perturbative analysis.\footnote{It is easy to check what happens when the second term in \eqref{eq:solutionofESBH2.5} is omitted: while for large masses the corrections to area and volume neglected in this way are comparably tiny, for small masses they are equal to the respective ``leading order'' terms, so one would underestimate both volume and entropy by a factor of two.} 
The scale parameter in \eqref{eq:action} in the notation of \cite{Grumiller:2005sq} is given by $\la=b\in\mathbb{R}^+$. In the asymptotic region $X\to\infty$ (``weak coupling'') the ESBH is well approximated by the Witten BH \cite{Witten:1991yr}, while near the origin $X\to 0$ (``strong coupling'') it approaches the Jackiw-Teitelboim model \cite{JT}. Because at the horizon the relation $\ga_h=2\sqrt{M(M-1)}$ holds, the BH area reads
\begin{equation}
  \label{eq:ESBHarea}
  A_{\rm ESBH}=N\left(2\sqrt{M(M-1)}+\arcsinh{(2\sqrt{M(M-1)})}\right)\,.
\end{equation}
Insertion into the volume definition \eqref{eq:vol1.5} and choosing $c$ conveniently yields
\begin{equation}
  \label{eq:ESBHvolume}
  V_{\rm ESBH}=N \left(2(M-1)+\frac{1}{2}\arcsinh^2{(2\sqrt{M(M-1)})}\right)\,.
\end{equation}
The mass parameter $M$ is bounded from below, $M\in[1,\infty)$. For strings in 2D one obtains $M=9/8$, but like in Refs.~\cite{Dijkgraaf:1992ba,Grumiller:2005sq} arbitrary values of $M$ are considered. As announced above, \eqref{eq:ESBHvolume} remains finite in the whole range of definition.

\begin{figure}
\epsfig{file=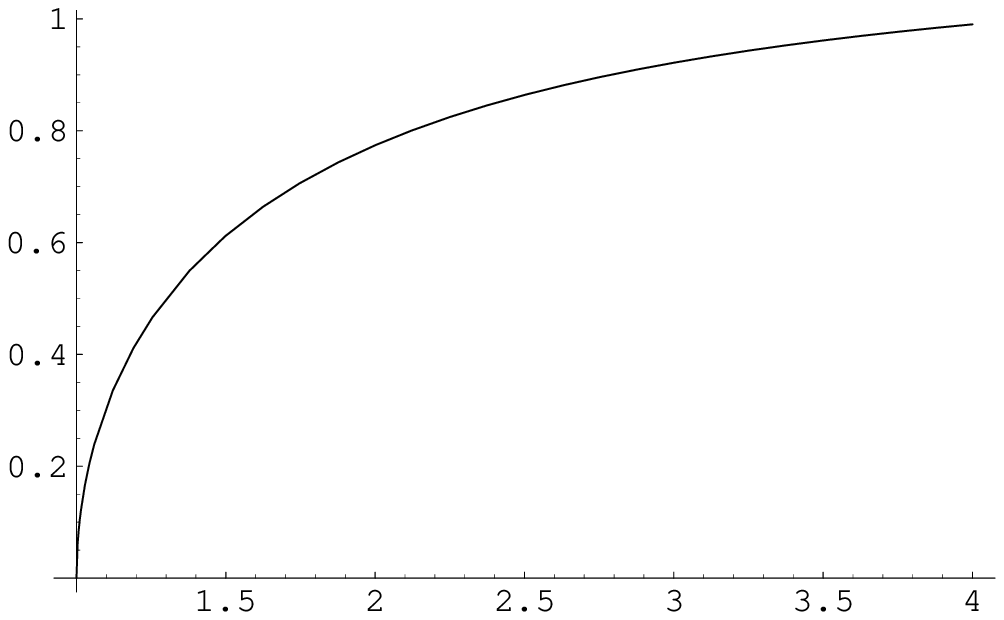,width=0.48\linewidth}\hspace{0.04\linewidth}
\epsfig{file=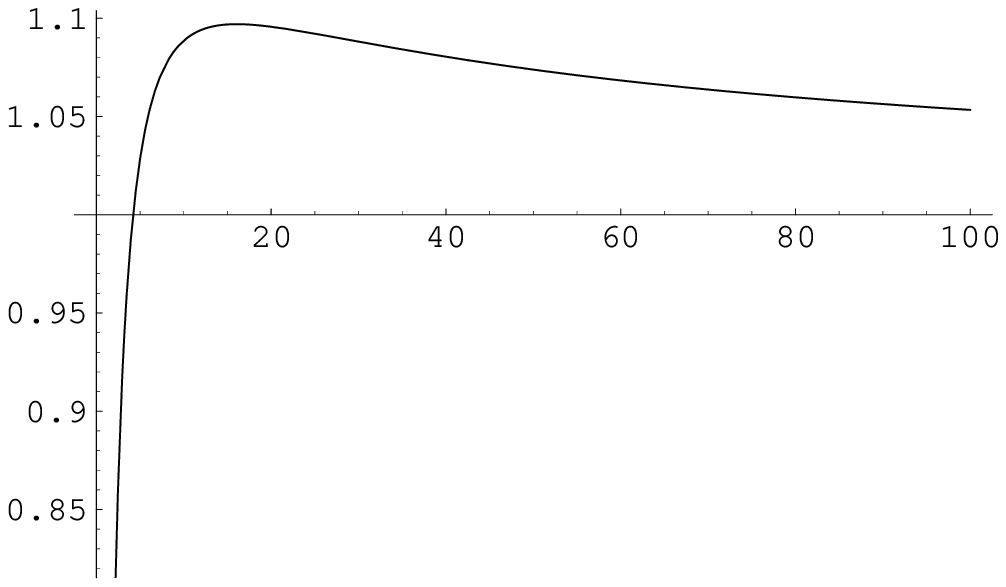,width=0.48\linewidth}
\caption{The ratio $V_{ESBH}/A_{ESBH}$ as function of $M$. The left (right) plot considers the range $M\in[1,4]$ ($M\in[1,100]$).}
\label{fig:2}
\end{figure}

For large masses the ratio $V_{\rm ESBH}/A_{\rm ESBH}$ approaches unity, concurrent with Eq.~\eqref{eq:vol11}. For small masses, $M=1+\eps$, the ratio $V_{\rm ESBH}/A_{\rm ESBH}\propto A_{\rm ESBH}\propto \sqrt{\eps}$ vanishes; the quadratic scaling $V_{\rm ESBH}\propto A^2_{\rm ESBH}$ in the strong coupling limit is expected because the Jackiw-Teitelboim model also exhibits such a behaviour. The ratio $V_{\rm ESBH}/A_{\rm ESBH}$ is plotted in Fig.~\ref{fig:2}. 
For large BHs the volume is always slightly larger (by $\ln^2{M}/(4M)$) than the area, while for very small BHs this trend is reversed until the ratio approaches zero. 

One may interpret these results as follows: for very large masses the error committed by scaling entropy with the volume (as implied by naive quantum field theoretical considerations) instead of scaling it with the area is almost negligible; there are only logarithmic corrections to the entropy arising in this manner. This is a remarkable property of the ESBH which it shares with the Witten BH. As expected, the ``quantum field theoretical entropy'' is always larger than the Bekenstein-Hawking entropy, but this feature is far less pronounced than, say, for the SBH where $V_{\rm SBH}\propto M^3$ and $A_{\rm SBH}\propto M^2$. While the ESBH evaporates down to $M=3$ the ratio $V_{\rm ESBH}/A_{\rm ESBH}$ deviates from its asymptotic value by less than $10\%$. Only for very small masses, $M\to 1$, the ``quantum field theoretical entropy'' actually underestimates the BH entropy appreciably. %In a sense, there are more degrees of freedom on the horizon than inside the BH.  
 
%\bibliographystyle{../fullsort}
%\bibliography{../review}

\begin{thebibliography}{9}

\bibitem{Parikh:2005}
Parikh M 2005 The Volume of Black Holes {\em Preprint}
  \href{http://www.arXiv.org/abs/hep-th/0508108}{{hep-th/0508108}}

\bibitem{Grumiller:2002nm} There is a lot of literature on 2D dilaton gravity. For sake of brevity only few references are provided. I apologize in advance to those whose work has not been included here. For a first orientation and a comprehensive list of references the following review may be consulted:

\item[]
Grumiller D, Kummer W and Vassilevich D V 2002 {\em Phys. Rept.} {\bf 369} 327 ({\em Preprint} \href{http://arXiv.org/abs/hep-th/0204253}{{hep-th/0204253}})
%%CITATION = HEP-TH 0204253;%%.

\bibitem{Klosch:1995fi}
Klosch T and Strobl T 1996
{\em Class. Quant. Grav.}  {\bf 13} 965 ({\em Preprint}
\href{http://arXiv.org/abs/gr-qc/9508020}{{gr-qc/9508020}})

\item[]
[Klosch T and Strobl T 1997 {\em Erratum-ibid.} {\bf 14} 825]

%%CITATION = GR-QC 9508020;%%

\bibitem{Gegenberg:1995pv}
Gegenberg J, Kunstatter G and Louis-Martinez D 1995 
{\em Phys. Rev.} {\bf D51} 1781 ({\em Preprint}
\href{http://arXiv.org/abs/gr-qc/9408015}{{gr-qc/9408015}})
%%CITATION = GR-QC 9408015;%%.

\bibitem{Katanaev:1997ni}
Katanaev M O, Kummer W and Liebl H 1997
{\em Nucl. Phys.} {\bf B486} 353 ({\em Preprint}
\href{http://www.arXiv.org/abs/gr-qc/9602040}{{gr-qc/9602040}})
%%CITATION = GR-QC 9602040;%%.

\bibitem{JT}
Teitelboim C 1983 
{\em Phys. Lett.} {\bf B126} 41

\item[]
Jackiw R 1985
{\em Nucl. Phys.} {\bf B252} 343
%%CITATION = NUPHA,B252,343;%%.


\bibitem{Witten:1991yr}
Witten E 1991
{\em Phys. Rev.} {\bf D44} 314
%%CITATION = PHRVA,D44,314;%%.
cf.~also the related CGHS model:

\item[]
Callan Jr. C G, Giddings S B, Harvey J A and Strominger A 1992
{\em Phys. Rev.} {\bf D45} 1005 ({\em Preprint}
\href{http://www.arXiv.org/abs/hep-th/9111056}{{hep-th/9111056}})
%%CITATION = HEP-TH 9111056;%%.

\bibitem{Dijkgraaf:1992ba}
Dijkgraaf R, Verlinde H and Verlinde E 1992
{\em Nucl. Phys.} {\bf B371} 269
%%CITATION = NUPHA,B371,269;%%.

\bibitem{Grumiller:2005sq}
Grumiller D 2005
{\em J. High Energy Phys.} JHEP05(2005)028 ({\em Preprint}
\href{http://www.arXiv.org/abs/hep-th/0501208}{{hep-th/0501208}})
%%CITATION = HEP-TH 0501208;%%.
cf.~also: 

\item[]
Grumiller D 2005 Logarithmic corrections to the entropy of the exact string black hole {\em Preprint} \href{http://www.arXiv.org/abs/hep-th/0506175}{{hep-th/0506175}}
%%CITATION = HEP-TH 0506175;%%


\end{thebibliography}

\section*{References}

\providecommand{\href}[2]{#2}\begingroup\raggedright\endgroup

\end{document}